\documentclass[aps,prd,amssymb,superscriptaddress,floatfix,nofootinbib,10pt]{revtex4}
\usepackage{times,amsbsy,amsmath,amsfonts,graphicx,float}
\usepackage{color,morefloats,rotating,srcltx,slashed}
\usepackage{multirow,bm,verbatim,tabularx,bbding,threeparttable}
\definecolor{dblue}{rgb}{0.00,0.00,0.75}
\usepackage[colorlinks,urlcolor=dblue,linkcolor=dblue,citecolor=dblue]{hyperref} 
\allowdisplaybreaks[4]

\newcommand{\PreserveBackslash}[1]{\let\temp=\\#1\let\\=\temp}
\newcolumntype{C}[1]{>{\PreserveBackslash\centering}p{#1}}
\newcolumntype{R}[1]{>{\PreserveBackslash\raggedleft}p{#1}}
\newcolumntype{L}[1]{>{\PreserveBackslash\raggedright}p{#1}}

\begin{document}

\title{Pentaquark molecular states with hidden bottom and double strangeness}

\begin{abstract}
We investigate the  meson-baryon interaction in coupled channels with the quantum numbers of the pentaquarks $P_{bss}$ and $P_{bsss}$. The interaction is derived from an extension of the local hidden gauge approach to the heavy quark sector, which has demonstrated accurate results compared to experiments involving $\Omega_{b}$, $\Xi_{b}$ states, and pentaquarks $P_{c}$ and $P_{cs}$.
In our study, we identify several molecular states with small decay widths within the chosen set of coupled channels. The spin-parity ($J^P$) of these states is as follows: $J^P={\frac{1}{2}}^-$ for pseudoscalar-baryon (${\frac{1}{2}}^+$) coupled channels, $J^P={\frac{3}{2}}^-$ for pseudoscalar-baryon (${\frac{3}{2}}^+$) coupled channels, $J^P={\frac{1}{2}}^-$ and ${\frac{3}{2}}^-$ for vector-baryon (${\frac{1}{2}}^+$) coupled channels, and $J^P={\frac{1}{2}}^-$, ${\frac{3}{2}}^-$, ${\frac{5}{2}}^-$ for vector-baryon (${\frac{3}{2}}^+$) coupled channels.
We search for the poles of these states and evaluate their couplings to the different channels. 
\end{abstract}


\date{\today}

\author{Jing Song}
\email[]{Song-Jing@buaa.edu.cn}
\affiliation{School of Physics, Beihang University, Beijing, 102206, China}
\affiliation{Departamento de Física Teórica and IFIC, Centro Mixto Universidad de Valencia-CSIC Institutos de Investigación de Paterna, 46071 Valencia, Spain}

\author{Man-Yu Duan }
\email[]{duanmy@seu.edu.cn}
\affiliation{Departamento de Física Teórica and IFIC, Centro Mixto Universidad de Valencia-CSIC Institutos de Investigación de Paterna, 46071 Valencia, Spain}
\affiliation{School of Physics, Southeast University, Nanjing 210094, China}

\author{ Luis Roca}
\email[]{luisroca@um.es}
\affiliation{Departamento de Física, Universidad de Murcia, E-30100 Murcia, Spain}

\author{ Eulogio Oset}
\email[]{oset@ific.uv.es}
\affiliation{Departamento de Física Teórica and IFIC, Centro Mixto Universidad de Valencia-CSIC Institutos de Investigación de Paterna, 46071 Valencia, Spain}

\maketitle

\section{Introduction}
The collaborative efforts at contemporary hadron facilities are leading to the discovery of numerous new states with heavy quarks, some of which are evidently exotic and deviate from the standard quark model of $q \bar q$ for mesons and $qqq$ for baryons. Specifically, there have been notable observations of baryons such as $\Lambda_c$~\cite{CLEO:1994oxm,ARGUS:1993vtm,CLEO:2000mbh,LHCb:2017jym,BaBar:2006itc}, $\Sigma_c$~\cite{Belle:2004zjl}, $\Xi_c$~\cite{CLEO:2000ibb,CLEO:1999msf,BaBar:2007xtc,LHCb:2020iby,Belle:2006edu,Belle:2020tom,BaBar:2007zjt}, $\Omega_c$~\cite{LHCb:2017uwr,LHCb:2023sxp}, $\Lambda_b$~\cite{LHCb:2012kxf,CDF:2013pvu,CMS:2020zzv,LHCb:2020lzx,LHCb:2019soc}, $\Sigma_b$~\cite{LHCb:2018haf}, $\Xi_b$~\cite{CMS:2021rvl,LHCb:2018vuc,LHCb:2020xpu,LHCb:2021ssn}, and $\Omega_b$~\cite{LHCb:2020tqd} (for a recent review of experimental findings, see Ref.~\cite{Chen:2022asf}). The discovery of hidden charm pentaquarks $P_c$~\cite{LHCb:2015yax,LHCb:2019kea,LHCb:2021chn} and hidden charm pentaquarks with strangeness $P_{cs}$~\cite{LHCb:2020jpq} has added significant excitement to the field. The search for new states is ongoing, and motivated by LHCb's plans to measure these new states, we focus here on the theoretical study of the $P_{bss}$ and $P_{bsss}$.

The $P_{c}$ and $P_{cs}$ states have attracted significant attention in past research through various theoretical models. Quark models have been extensively employed to study these states~\cite{Gershtein:1998sx,Gershtein:2000nx,Ebert:2002ig,Kiselev:2002iy,Roberts:2007ni,Valcarce:2008dr,Albertus:2009ww,Albertus:2012jt,Ma:2015lba,Shah:2016vmd,Garcilazo:2016piq,Lu:2017meb,Xiao:2017dly,Niu:2018ycb,Salehi:2018oky,Li:2019ekr,Gutierrez-Guerrero:2019uwa,Shah:2021reh,Ghalenovi:2022dok}. Furthermore, they have been investigated using lattice QCD techniques~\cite{Garcilazo:2016piq,Lewis:2001iz,Lin:2011ti,Brown:2014ena,Perez-Rubio:2015zqb,Bahtiyar:2018vub,Briceno:2012wt} and QCD sum rules~\cite{Wang:2010hs,Hu:2017dzi,Wang:2018lhz}. Additional methods have also been applied in other studies~\cite{Weng:2010rb,Karliner:2015ina,Soto:2020pfa,Soto:2021cgk}.

Research on meson-baryon interactions is extensive, as reviewed in Refs.~\cite{Oller:2000ma,Guo:2017jvc}. A successful approach involves exchanging vector mesons to predict various baryonic states with open charm or bottom quarks. For instance, the $\Xi_{cc}$ states were examined in Ref.~\cite{Dias:2018qhp}, and this was extended to $\Xi_{bb}$ and $\Omega_{bbb}$ states in Ref.~\cite{Dias:2019klk}. Studies on $\Xi_c$ and $\Xi_b$ states were conducted in Ref.~\cite{Yu:2018yxl}, $\Xi_{bc}$ states in Ref.~\cite{Yu:2019yfr}, and $\Omega_b$ states in Ref.~\cite{Liang:2017ejq}. Some of these predicted states match recently discovered experimental states~\cite{Chen:2022asf}. In particular, the predicted $\Omega_b$ states in Ref.~\cite{Liang:2017ejq} might correspond to observations reported in Ref.~\cite{LHCb:2020tqd} as shown in Ref.~\cite{Liang:2020dxr}. Each study involves detailed analysis of interactions among multiple coupled channels.

Other studies also explore these states from a molecular perspective, using different methods and dynamics. For instance, in Ref.~\cite{Shimizu:2017xrg}, researchers used one-pion exchange and $D^{(*)}$ exchange to study the dynamics of $\Xi_{cc}$ states. in Ref.~\cite{Guo:2013xga}, heavy flavor, heavy quark spin, and heavy antiquark-diquark symmetries are applied within an effective field theory framework to study pentaquarks and open bottom baryonic states. The one-boson exchange model is used to investigate $\Omega_c$ states in Ref.~\cite{Chen:2017xat}.

Moreover, extensive research has been conducted on molecular states arising from meson-baryon interactions in different sectors—light, charm, and bottom—using $SU(6){\mathrm{lsf}} \times SU(2){\mathrm{HQSS}}$ symmetry. This involves using $SU(6)$ flavor-spin symmetry in the light sector and $SU(2)$ in the heavy sector, while respecting heavy quark spin symmetry~\cite{Romanets:2012hm,Garcia-Recio:2013gaa,Garcia-Recio:2012lts}. These studies extend the dynamics from the Weinberg-Tomozawa interaction and are noteworthy for their ability to correlate multiple coupled channels across different sectors. They make qualitative predictions for bound states and resonances with a wide range of quantum numbers. Recently, this approach has been used to address $\Xi_b$ and $\Xi_c$ states in Ref.~\cite{Nieves:2019jhp}.

Our goal is to investigate the states that can form as molecular states from the s-wave interaction of mesons with baryons in their ground state. Consequently, we expect to find only baryon states with negative parity, which differ from most states predicted by quark models. The attractive force between mesons and baryons often makes the formation of these states inevitable. Additionally, the proximity of the masses of some states to the threshold of certain meson-baryon channels necessitates the explicit consideration of these channels and their interactions in any comprehensive study of the baryon spectrum~\cite{Dong:2021rpi}.

Many studies on molecular states in the charm sector have focused on the pentaquarks $P_c$, $P_{cs}$, and $P_{css}$. These states contain hidden charm, and have been reviewed in a number of papers~\cite{Dong:2021bvy,Dong:2021juy,Guo:2017jvc,Chen:2016qju,Dong:2017gaw,Liu:2019zoy,Ramos:2020bgs,Wang:2020bjt,Marse-Valera:2022khy,Roca:2024nsi}, that also cover open heavy quark baryonic molecular states.

In the present paper we study baryon states with hidden bottom and two or three $s$ quarks. The precursor of these states are the $P_{css}$, $P_{csss}$. The $P_{css}$ states were studied from the molecular perspective, as meson baryon states in Refs.~\cite{Wang:2020bjt,Marse-Valera:2022khy,Roca:2024nsi}. 
In Ref.~\cite{Marse-Valera:2022khy} a pole was found for states of pseudoscalar-baryon($1/2^+$) nature ($PB$) and another one for states of vector-baryon($1/2^+$) nature ($VB$). In Ref.~\cite{Roca:2024nsi} such states were also found and, in addition, two more states of $PB^*$ and $VB^*$ nature were found with $B^*$ corresponding to baryons of $3/2^+$. The $P_{csss}$, of $c\bar{c}sss$ nature was not found as a bound state. 
Similar results were found in Ref.~\cite{Wang:2020bjt}, where, however, under certain circumstance the $P_{csss}$ state could be also formed. It is important to mention that the dynamics of coupled channels was found essential to obtain the bound states, that were not generated in Refs.~\cite{Wu:2010vk,Xiao:2013yca}, where some important non diagonal channels were neglected. 

In the present  work we retake this line of research and look for possible states of $b \bar{b} ssq$ and $b \bar{b} sss$ nature. The only free parameter of the theory is constrained by the experimental results on the $P_{cs}$~\cite{LHCb:2020jpq} and $\Omega_c$~\cite{LHCb:2017uwr,LHCb:2023sxp} states that were studied along the same lines in Refs.~\cite{Debastiani:2017ewu,Feijoo:2022rxf}, which should lead to reliable results in the predictions of  $b \bar{b} ssq$ and $b \bar{b} sss$ involving arguments of heavy quark symmetry.


\section{Formalism}


We consider meson-baryon coupled channels: $PB$,  $VB$, $PB^*$, and $VB^*$, where $P$ represents the pseudoscalar meson, $V$ denotes vector meson, $B$ refers to ground state baryons with $J^P={\frac{1}{2}}^+$, and $B^*$ indicates ground state baryons with $J^P={\frac{3}{2}}^+$. We do not mix these states as justified in Ref.~\cite{Roca:2024nsi} and discussed below.
\begin{itemize}
  \item [i)] $PB$ channels:
                $\eta_b\Xi^0$, ${B}^{0}_{s}\Xi^{'}_b$, and $B^+\Omega_b^-$.

  \item [ii)] $VB$ channels:
                $\Upsilon\Xi^0$, ${B}^{*0}_{s}\Xi^{'}_b$, and $B^{*+}\Omega_b^-$.
 
  \item [iii)] $PB^*$ channels:
                 $\eta_b\Xi^{*0}$, ${B}^{0}_{s}\Xi^{*0}_b$, and $B^+\Omega_b^{*-}$.
  
  \item [iv)] $VB^*$ channels:
                 $\Upsilon\Xi^{*0}$, ${B}^{*0}_{s}\Xi^{*0}_b$, and $B^{*+}\Omega_b^{*-}$.   
  \item [v)]  $P B^*: \eta_b \Omega^-, B_s^0 \Omega_b^{*-}$
  \item [vi)]   $V B^*: \Upsilon \Omega^-, B_s^{*0} \Omega_b^{*-}$
\end{itemize}

The first four sectors correspond to $b \bar{b} ssq$ states and the last two to $b \bar{b} sss$.

The interaction model we use is based on exchanging vector mesons, adapted from the local hidden gauge approach~\cite{Bando:1984ej,bando1988nonlinear,Harada:2003jx,Meissner:1987ge,Nagahiro:2008cv} to apply to the bottom sector~\cite{Dias:2019klk,Yu:2018yxl,Yu:2019yfr,Liang:2017ejq,Dong:2021bvy,Dong:2021juy,Kong:2021ohg}. To mix the interaction blocks, pion  or other pseudoscalar exchanges are needed, but these contributions are much less significant than vector meson exchange when it comes to determining the masses of the states. While pion exchange can influence the widths of these states, in many cases where multiple channels are involved and decay to lower mass states can happen via vector exchange, the effect of pion exchange is still relatively small (see the appendix of Ref.~\cite{Dias:2021upl}). We focus on interactions in the $S$-wave, which helps define the $J^P$ (spin-parity) characteristics of the states, and the relevant coupled channels are listed below, together with their threshold masses.
\begin{figure}[H]   
  \centering
  \includegraphics[width=7.5cm]{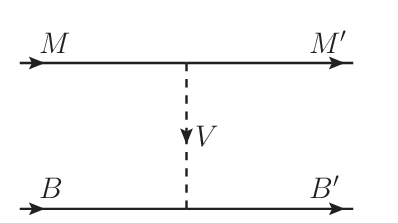}
  \caption{The interaction between $MB\to M^\prime B^\prime$ happens through the exchange of vector mesons. Here $M (M^\prime)$ are the initial and final mesons, and  $B (B^\prime)$ are the initial and final baryons. The vector meson involved in this exchange is represented by $V$.
              }
  \label{fig1}
\end{figure}

\begin{table}[H]
\centering
\caption{Threshold masses (in MeV) of different channels.}
\label{thres}
\setlength{\tabcolsep}{38pt}
\begin{tabular}{l|cccc}
\hline \hline
  \multirow{2}*{$PB,~J^P={\frac{1}{2}}^-$} 
&$\eta_b\Xi^0$ & ${B}^{0}_{s}\Xi^{'}_b$ & $ B^+\Omega_b^-$ \\
~ & 10713.6 & 11301.9 & 11324.5 \\
\hline
  \multirow{2}*{$VB,~J^P={\frac{1}{2}}^-,{\frac{3}{2}}^-$}
                         & $\Upsilon\Xi^0$ & ${B}^{*0}_{s}\Xi^{'}_b$ & $B^{*+}\Omega_b^-$ \\
~ & 10775.3 & 11350.4 & 11369.9 \\
\hline
  \multirow{2}*{$PB^*, ~J^P={\frac{3}{2}}^-$} 
                        & $\eta_b\Xi^{*0}$ & ${B}^{0}_{s}\Xi^{*0}_b$ & $B^+\Omega_b^{*-}$ \\
~ & 10930.5 & 11319.2 & 11381.3 \\

\hline
  \multirow{2}*{$VB^*,~J^P={\frac{1}{2}}^-,{\frac{3}{2}}^-,{\frac{5}{2}}^-$}
                        & $\Upsilon\Xi^{*0}$ & ${B}^{*0}_{s}\Xi^{*0}_b$ & $B^{*+}\Omega_b^{*-}$ \\
~ & 10992.2 & 11367.7 & 11426.7  \\  
\hline
  \multirow{2}*{$PB^*, ~J^P={\frac{3}{2}}^-$} 
                        & $  \eta_b \Omega^-$ & $B_s^0 \Omega_b^{*-}$ \\
~ &  11071.2 & 11468.9  \\

\hline
  \multirow{2}*{$VB^*,~J^P={\frac{1}{2}}^-,{\frac{3}{2}}^-,{\frac{5}{2}}^-$}
                        & $ \Upsilon \Omega^-$ & $B_s^{*0} \Omega_b^{*-}$ \\
~ & 11132.9  & 11517.4  \\

\hline
\hline
\end{tabular}
\end{table}

As mentioned earlier, we use vector exchange based on an extended local hidden gauge approach between mesons and baryons, as shown in Fig.~\ref{fig1}. For our set of states, there are two types of $VMM^\prime$ vertices: $VPP$ (where $V$ stands for vector and $P$ stands for pseudoscalar) and $VVV$. These are described by the following Lagrangians~\cite{Nagahiro:2008cv}:
\begin{align}
    \mathcal{L}_{\mathrm{VPP}} &= -i g\left\langle\left[P, \partial_{\mu} P\right] V^{\mu}\right\rangle, \label{eq-1}\\
    \mathcal{L}_{\mathrm{VVV}} &=  i g\left\langle\left(V^{\mu} \partial_{\nu} V_{\mu}-\partial_{\nu} V^{\mu}
                                     V_{\mu} \right) V^{\nu}\right\rangle. \label{eq-2}    
\end{align}
The coupling parameter $g$ is determined by $g = \frac{m_V}{2f_\pi}$, where $m_V = 800$ MeV represents the vector meson mass and $f_\pi = 93$ MeV denotes the pion decay constant. In this context, $P$ or $V$ refer to the $q_i\bar{q}_j$ matrices expressed in terms of mesons. The symbol $\langle\cdot \cdot \cdot\rangle$ signifies the trace operation for these matrices. It is important to note that although $q_i\bar{q}_j$ matrices belong to the SU(4) or SU(5) groups, the vertices described in Eqs.~(\ref{eq-1}) and (\ref{eq-2}) only involve the overlap of $q\bar{q}$ in the external mesons and the exchanged vectors. Therefore, the utilization of SU(4) or SU(5) symmetry is redundant, as discussed in Ref.~\cite{Sakai:2017avl}.

The matrices $P$ and $V$ that we need are given by, 
\begin{equation}
 \label{eq:matP_bottom}
    P = \begin{pmatrix}
            \frac{1}{\sqrt{2}}\pi^0 + \frac{1}{\sqrt{3}} \eta + \frac{1}{\sqrt{6}}\eta' & \pi^+ & K^+ & B^+ \\
            \pi^- & -\frac{1}{\sqrt{2}}\pi^0 + \frac{1}{\sqrt{3}} \eta + \frac{1}{\sqrt{6}}\eta' & K^0 & B^0 \\
            K^- & \bar{K}^0 & -\frac{1}{\sqrt{3}} \eta + \sqrt{\frac{2}{3}}\eta' & B_s^0 \\
            B^-  & \bar B^0 & \bar B_s^0 & \eta_b
         \end{pmatrix},
\end{equation}
\begin{equation}
\label{eq:matV_bottom}
   V = \begin{pmatrix}
            \frac{1}{\sqrt{2}}\rho^0 + \frac{1}{\sqrt{2}} \omega & \rho^+ & K^{* +} & B^{* +} \\
            \rho^- & -\frac{1}{\sqrt{2}}\rho^0 + \frac{1}{\sqrt{2}} \omega  & K^{* 0} & B^{* 0} \\
           K^{* -} & \bar{K}^{* 0}  & \phi & B_s^{* 0} \\
           B^{* -} & \bar B^{* 0} & \bar B_s^{* 0} & \Upsilon
         \end{pmatrix}.
\end{equation}
Only the terms of $B,~B_s$ pseudoscalars and $\eta_b$ of the $P$ matrix enter the present calculation.
The interaction obtained for the mechanism of Fig.~\ref{fig1}  is always of the type
\begin{eqnarray}
  \label{eq:def_Vij}
     V_{ij}= g^2(p_1^0+p^{0}_3)C_{ij} \, ,
\end{eqnarray}
with $g=\frac{M_V}{2f_\pi}$, $M_V=800~\mathrm{MeV},~f_\pi=93$~MeV, where $p_1^0$, $p^{0}_3$ are the energies of the initial and final mesons, respectively. The coefficient $C_{ij}$ are then evaluated 
and we find the following tables (Tables~\ref{PotentialA},~\ref{PotentialB},~\ref{PotentialC},~\ref{PotentialD}) for these interactions.

\begin{table}[H]
\centering
 \caption{ Coefficients $C_{ij}$ for the PB sector with $J^P = \frac{1}{2}^{-}$.}\label{PotentialA}
\setlength{\tabcolsep}{40pt}
\begin{tabular}{cccccccc}
\hline
\hline
States & $\eta_b\Xi^0$ & ${B}^{0}_{s}\Xi^{'}_b$ & $B^+\Omega_b^-$ \\
\hline
$\eta_b\Xi^0$                & 0 & $\frac{1}{\sqrt{6}}~\frac{1}{M_{B_s^*}^2}$ & $\frac{-1}{\sqrt{3}}~\frac{1}{M_{B^*}^2}$ \\
 ${B}^{0}_{s}\Xi^{'}_b$  &   & $\frac{1}{M_\phi^2}-\frac{1}{M_{\Upsilon}^2}$ & $\frac{\sqrt{2}}{M_{K^{*}}^2}$ \\
$B^+\Omega_b^-$      &   &                                             & $-\frac{1}{M_{\Upsilon}^2}$ \\
\hline
\hline
\end{tabular}
\end{table}

\begin{table}[H]
\centering
 \caption{ Coefficients $C_{ij}$ for the VB sector with $J^P = \frac{1}{2}^{-}, \frac{3}{2}^{-}$.}\label{PotentialB}
\setlength{\tabcolsep}{40pt}
\begin{tabular}{cccccccc}
\hline
\hline
States & $\Upsilon\Xi^0$ & ${B}^{*0}_{s}\Xi^{'}_b$ & $B^{*+}\Omega_c^0$ \\
\hline
$\Upsilon\Xi^0$                 & 0 & $\frac{1}{\sqrt{6}}~\frac{1}{M_{B_s^*}^2}$ & $\frac{-1}{\sqrt{3}}~\frac{1}{M_{B^*}^2}$ \\
${B}^{*0}_{s}\Xi^{'}_b$  &   & $\frac{1}{M_\phi^2}-\frac{1}{M_{\Upsilon}^2}$ & $\frac{\sqrt{2}}{M_{K^{*}}^2}$ \\
$B^{*+}\Omega_c^0$      &   &                                             & $-\frac{1}{M_{\Upsilon}^2}$ \\
\hline
\hline
\end{tabular}
\end{table}

\begin{table}[H]
\centering
 \caption{ Coefficients $C_{ij}$ for the PB sector with $J^P = \frac{3}{2}^{-}$.}\label{PotentialC}
\setlength{\tabcolsep}{40pt}
\begin{tabular}{cccccccc}
\hline
\hline
States & $\eta_b\Xi^{*0}$ & ${B}^{0}_{s}\Xi^{*0}_b$ & $B^+\Omega_b^{*-}$ \\
\hline
$\eta_b\Xi^{*0}$                &  0 & $\frac{\sqrt{2}}{\sqrt{3}}~\frac{1}{M_{B_s^*}^2}$ & $\frac{1}{\sqrt{3}}~\frac{1}{M_{B^*}^2}$  \\
${B}^{0}_{s}\Xi^{*0}_b$     &   & $\frac{1}{M_\phi^2}-\frac{1}{M_{\Upsilon}^2}$ & $\frac{\sqrt{2}}{M_{K^{*}}^2}$ \\
$B^+\Omega_b^{*-}$      &   &                                             & $-\frac{1}{M_{\Upsilon}^2}$ \\
\hline
\hline
\end{tabular}
\end{table}

\begin{table}[H]
\centering
 \caption{ Coefficients $C_{ij}$ for the VB sector with $J^P = \frac{1}{2}^{-}, \frac{3}{2}^{-},  \frac{5}{2}^{-}$.}\label{PotentialD}
\setlength{\tabcolsep}{40pt}
\begin{tabular}{cccccccc}
\hline
\hline
States & $\Upsilon\Xi^{*0}$ & ${B}^{*0}_{s}\Xi^{*0}_b$ & $B^{*+}\Omega_b^{*-}$ \\
\hline
$\Upsilon\Xi^{*0}$                  &  0 & $\frac{\sqrt{2}}{\sqrt{3}}~\frac{1}{M_{B_s^*}^2}$ & $\frac{1}{\sqrt{3}}~\frac{1}{M_{B^*}^2}$  \\
${B}^{*0}_{s}\Xi^{*0}_b$      &   & $\frac{1}{M_\phi^2}-\frac{1}{M_{\Upsilon}^2}$ & $\frac{\sqrt{2}}{M_{K^{*}}^2}$ \\
$B^{*+}\Omega_b^{*-}$       &   &                                             & $-\frac{1}{M_{\Upsilon}^2}$ \\
\hline
\hline
\end{tabular}
\end{table}

\begin{table}[H]
\centering
 \caption{ Coefficients $C_{ij}$ for the $PB^*$ sector with $J^P = \frac{3}{2}^{+}, \frac{3}{2}^{-}$.}\label{PotentialE}
\setlength{\tabcolsep}{60pt}
\begin{tabular}{cccccccc}
\hline
\hline
States & $  \eta_b \Omega^-$ & $B_s^0 \Omega_b^{*-}$ \\
\hline
$\eta_b \Omega^-$                  &  0 & $\frac{1}{M_{B_s^*}^2}$  \\
$B_s^0 \Omega_b^{*-}$      &   & $\frac{2}{M_\phi^2}-\frac{1}{M_{\Upsilon}^2}$\\
\hline
\hline
\end{tabular}
\end{table}

\begin{table}[H]
\centering
 \caption{ Coefficients $C_{ij}$ for the $VB^*$ sector with $J^P = \frac{3}{2}^{+}, \frac{3}{2}^{-},  \frac{5}{2}^{-}$.}\label{PotentialE}
\setlength{\tabcolsep}{60pt}
\begin{tabular}{cccccccc}
\hline
\hline
States & $ \Upsilon \Omega^-$ & $B_s^{*0} \Omega_b^{*-}$  \\
\hline
$ \Upsilon \Omega^-$                  &  0 & $\frac{1}{M_{B_s^*}^2}$  \\
$B_s^{*0} \Omega_b^{*-}$      &   & $\frac{2}{M_\phi^2}-\frac{1}{M_{\Upsilon}^2}$\\
\hline
\hline
\end{tabular}
\end{table}   

After figuring out the $V_{ij}$ potential, we use the Bethe-Salpeter equation expressed in matrix form to obtain the scattering matrix for the channels involved
\begin{eqnarray}
   T = [1-VG]^{-1}V,
\end{eqnarray}
where $G$ represents a diagonal matrix containing the loop functions for the intermediate states of meson and baryon. We adopt the cutoff approach to regularize the loops as outlined in Ref.~\cite{Roca:2024nsi}, setting $q_\mathrm{max}=600$ MeV. $G_l$ denotes the meson-baryon loop function\footnote{For some channels far below the energy under consideration, it is common to consider only $\mathrm{Im}~G$, which has repercussion on the widths, and neglect $\mathrm{Re}~G$~\cite{Dai:2020yfu}. We follow a slightly different  prescription taking $\mathrm{Re}~G$=0 when $\mathrm{Re}~G$ becomes bigger than 0 and use the exact $\mathrm{Im}~G$. }.

\begin{align}\label{cut}
   G_l(\sqrt{s})= \int_{|q|<q_\mathrm{max}} \frac{d^3 q}{(2 \pi)^3} \frac{2~M_l~(w_l(q)+E_l(q))}{2 w_l(q) E_l(q))} \frac{1}{s-(w_l(q)+E_l(q))^2+i \epsilon} ,
\end{align}
where $M_l$ stands for the mass of the baryon, $m_l$ for the meson and $w_l(q)=\sqrt{m_l^2+\Vec{q}~^2}$,  $E_l(q)=\sqrt{M_l^2+\Vec{q}~^2}$.
In the  meson baryon rest frame, the total incident momentum $P$ is $(\sqrt{s}, 0,0,0)$. In the region of interest, the real part of $G$ is negative.

Poles are encountered in the second Riemann sheet, necessitating a change from the loop function in the first sheet, $G^I$,  to that in the second one, $G^{II}$, as follows:
\begin{eqnarray}\label{G2}
G^{II}_j = G^I_j + i \frac{2M_j\,q}{4\pi\sqrt{s}}\,,
\end{eqnarray}
for Re$\sqrt s>m_j+M_j$, and $q$ given by
\begin{eqnarray}\label{qcm}
q = \frac{\lambda^{1/2}(s,m^2_j,M^2_j)}{2\sqrt{s}}\,,\qquad(\mathrm{Im}~q>0).
\end{eqnarray}
We further assess the couplings defined by the residue at the pole, where the amplitudes behave as:
\begin{equation}
T_{ij} = \frac{g_i g_j}{z-z_R} \, ,
\end{equation}
with $z_R$ representing the complex energy ($M, i\Gamma/2$). For one $g_i$, we select a sign, while the remaining couplings have their relative signs well defined. Additionally, we present $g_iG^{II}_i$, yielding the wave function at the origin in coordinate space~\cite{Gamermann:2009uq}.

Moreover, we can assess the compositeness, denoted as $X_i$, of a generated resonance in a given channel, $i$. This value indicates the weight of a molecular component in the state $i$ in the wave function, as described in Ref.~\cite{Gamermann:2009uq,Dai:2023kwv,Song:2023pdq}.
$$
\chi_i=-\left.g_i^2 \frac{\partial G_i}{\partial \sqrt{s}}\right|_{\sqrt{s}} .
$$

When searching for pole positions, employing $G_l^I(\sqrt{s})$ for $\operatorname{Re}(\sqrt{s})<m_l+M_l$ and $G_l^{II}(\sqrt{s})$ for $\operatorname{Re}(\sqrt{s})>m_l+M_l$ brings us closer to the pole positions and half-widths akin to those of the corresponding Breit-Wigner forms along the real axis. This method also aids in identifying potential pure bound states situated below the lowest threshold, where the prescription of Eq.~(\ref{G2}) with the $\mathrm{Re}\sqrt{s}>M_l+m_l$ renders us to the first Riemann sheet.

\section{results}
In Tables~\ref{res1},~\ref{res2},~\ref{res3},~\ref{res4} we show the results obtained using $q_\mathrm{max}=600$~MeV and $q_\mathrm{max}=800$~MeV.
\begin{table}[H]
\footnotesize
\centering
\caption{$PB$ states (pole position in units of {MeV}). The first row in poles, $g_i$ and $~\chi_i$ correspond to $q_\mathrm{max}=600$~MeV and the second one to $q_\mathrm{max}=800$~MeV. $\chi_i$ stands for compositeness. The last column is the sum of the different compositeness. }\label{res1} 
\setlength{\tabcolsep}{16pt}
\begin{tabular}{l|lcccc}
\hline \hline
    Poles           &       ~           & $\eta_b\Xi^0$ & ${B}^{0}_{s}\Xi^{'}_b$ & $ B^+\Omega_b^-$  \\
\hline
{$(11259.7-i~ 0.16) $} 
                        & \multirow{2}*{ $g_i$} 	&	$ ( 0.12+i~0.00    )                         $ & $ ( -2.62-i~ 0.01  ) $ & $(  4.03+i~0.01 )  $	\\
 $(11166.0-i~0.36 ) $                       &   	&	$ (  0.19-i~ 0.00    )                         $ & $ ( -4.30-i~0.00   ) $ & $(  5.98+i~0.01    )  $	\\
  & \multirow{2}*{ $\chi_i$}  	& $(  -0.00+i~0.00    )   $ & $ (  0.42-i~ 0.00 )  $ & $ (  0.57+i~0.00 )$  & $ (  0.99-i~ 0.00   )$	\\  
                        &   	&	$ ( -0.00+i~0.00     )                         $ & $ ( 0.38-i~0.00   ) $ & $( 0.60-i~ 
0.00   )  $ &	$ (  0.98-i~ 0.00  ) $	\\
\hline
\hline
\end{tabular}
\end{table}

\begin{table}[H]
\footnotesize
\centering
\caption{Same as Table~\ref{res1} for $VB$ states.}\label{res2} 
\setlength{\tabcolsep}{16pt}
\begin{tabular}{l|lcccc}
\hline \hline
    Poles           &       ~           & $\Upsilon\Xi^0$ & ${B}^{*0}_{s}\Xi^{'}_b$ & $B^{*+}\Omega_b^-$ \\
\hline
{$ (11306.2-i~ 0.16)    $} 
                        & \multirow{2}*{ $g_i$}  			&	$ (  0.12+i~0.00   )                          $ &  $( -2.67-i~ 0.01  )  $ & $ ( 4.02+i~0.01  )$   \\

$(11212.3-i~0.35 ) $                         &   	&	$ ( 0.19-i~0.00     )                         $ & $ ( -4.35-i~ 0.01   ) $ & $( 5.99+i~0.01   )  $	\\
            & \multirow{2}*{ $\chi_i$} 	& $( -0.00+i~0.00   )   $ & $ ( 0.41-i~ 0.00  )  $ & $ (  0.58+i~0.00  )$  & $ ( 0.99-i~ 0.00  )$	\\
                         &   	&	$ (  -0.00+i~0.00    )                         $ & $ (0.38-i~0.00   ) $ & $(  0.60-i~ 
0.00
  )  $ & $ ( 0.98-i~0.00   ) $	\\
\hline
\hline
\end{tabular}
\end{table}

\begin{table}[H]
\footnotesize
\centering
\caption{Same as Table~\ref{res1} for $PB^*$ states.}\label{res3} 
\setlength{\tabcolsep}{16pt}
\begin{tabular}{l|lcccc}
\hline \hline
    Poles           &       ~           & $\eta_b\Xi^{*0}$ & ${B}^{0}_{s}\Xi^{*0}_b$ & $B^+\Omega_b^{*-}$  \\
\hline
{$  (11297.8-i~ 0.01)  $} 
              ~          & \multirow{2}*{ $g_i$} 	&  $(  0.03+i~0.00   )    $ & $(    -1.99-i~ 0.00  ) $ & $ ( 4.35-i~ 
 0.00 )$	\\
$(11206.6-i~ 0.00) $                        &   	&	$ ( 0.02+i~0.00     )                         $ & $ ( -3.94-i~ 0.00   ) $ & $(  6.23-i~ 
 0.00
  )  $	\\
                      & \multirow{2}*{ $\chi_i$} 	& $( -0.00+i~0.00    )   $ & $ ( 0.52+i~0.00  )  $ & $ ( 0.47-i~ 
0.00  )$  & $ ( 0.99+i~0.00    )$	\\ 
                        &   	&	$ (  -0.00+i~0.00     )                         $ & $ ( 0.42+i~0.00 ) $ & $(    0.56-i~ 0.00
  )  $ & $ ( 0.98+i~0.00   ) $	\\
\hline
\hline
\end{tabular}
\end{table}

\begin{table}[H]
\footnotesize
\centering
\caption{Same as Table~\ref{res1} for $VB^*$ states.}\label{res4} 
\setlength{\tabcolsep}{16pt}
\begin{tabular}{l|lcccc}
\hline \hline
    Poles           &       ~           & $\Upsilon\Xi^{*0}$ & ${B}^{*0}_{s}\Xi^{*0}_b$ & $B^{*+}\Omega_b^{*-}$  \\
\hline
  {$  (11344.6-i~0.01)  $} 
              ~          & \multirow{2}*{ $g_i$} 	&  $(  0.03+i~0.00    )    $ & $(   -2.06-i~ 0.00   ) $ & $ ( 4.36-i~ 
0.00 )$	\\
    $( 11253.0-i~ 0.00) $                    &   	&	$ ( 0.02+i~0.00     )                         $ & $ ( -3.99-i~ 0.00   ) $ & $(  6.24-i~ 0.00  
  )  $	\\
                      & \multirow{2}*{ $\chi_i$} 	& $(  -0.00+i~0.00     )   $ & $ ( 0.51+i~0.00  )  $ & $ ( 0.48-i~ 0.00   )$  & $ ( 0.99+i~0.00     )$	\\ 
                        &   	&	$ ( -0.00+i~0.00    )                         $ & $ ( 0.41+i~0.00  ) $ & $( 0.57-i~0.00   
  )  $ & $ (  0.98+i~0.00  ) $	\\
\hline
\hline
\end{tabular}
\end{table}

As we can see in the Table~\ref{res1},  for the $PB$ channel, we get a bound state with extremely small width. The state couples mostly both to ${B}^{0}_{s}\Xi^{'}_b$ and $ B^+\Omega_b^-$, and for $q_\mathrm{max}=600$~MeV, it is bound by about 40~MeV with respect to the ${B}^{0}_{s}\Xi^{'}_b$ threshold and couples very weakly to the $\eta_b\Xi^0$, the only channel open for decay, hence the small width obtained. The wave functions at the origin follow the same trend as the couplings. Table~\ref{res2} shows results for the $VB$ states, very similar to those of the $PB$ case. The main couplings  are to ${B}^{*0}_{s}\Xi^{'}_b$ and $B^{*+}\Omega_b^-$, and the  $\Upsilon\Xi^0$ has a small coupling, leading to a very small width. The results are also similar in the $PB^*$ and $VB^*$ channels in Tables~\ref{res3},\ref{res4}.

We show also the results for $q_\mathrm{max}=800$~MeV to have a feeling for uncertainties, but the value of $q_\mathrm{max}=600$~MeV is closer to the one used to get the $\Omega_c$ in Ref.~\cite{Debastiani:2017ewu} and $P_{cs}$ states in Ref.~\cite{Feijoo:2022rxf}. According to~\cite{Aceti:2014ala,Lu:2014ina}, improving heavy quark symmetry demands the use of the same cut off in the charm and bottom sectors. The interesting thing is that in all four cases we get a bound state with very small width.

We give now some words about the probabilities found for each channels $\chi_i$ . The values of $\chi_i$ are complex, because we have small widths from the decay channels. In this case the proper interpretation of the meaning of $\chi_i$ can be seen in Ref.~\cite{Aceti:2014ala}.
However, given the very small imaginary parts, the interpretation of the real part as a measure of the probability to find the given component in the wave function is appropriate. We find the total probability close to 1 indicating the molecular component of the states, which is natural from their construction. For the  triple strangeness sector, but we do not find bound states, nor even cusps around thresholds. 

\section{Conclusions} 

  We have addressed the issue of possible pentaquark states of type $b \bar{b} ssq$ and $b \bar{b} sss$ of molecular nature, stemming from the interaction of mesons with baryons in their ground state. We have considered four types of such states coming from the interaction of $PB$, $VB$, $PB^*$, $VB^*$, where $P$ stands for pseudoscalar mesons, $V$ for vector mesons and $B$, $B^*$ for baryons of $1/2^+$ and $3/2^+$ respectively. We have made implicit use of heavy quark symmetry using cutoff regularization of the loops with the same cutoff in the bottom sector as in the charm sector~\cite{Lu:2014ina,Ozpineci:2013zas}. We have then relied upon the analysis of the $\Omega_c$ and $P_{cs}$ states along the same lines as in the present work to get reasonably values of this cutoff.  We obtain bound states of $b \bar{b} ssq$ type, with an extremely small width, for each of these four sectors, which are bound by about 50~MeV. We have also addressed the $b \bar{b} sss$ sector, but in this case we do not find any state of this type. We are looking forward to future experimental search of these states in LHCb or other facilities. The work presented here not only shows the existence of these states, but also the open channels where they decay, which should serve as guidance for future experiments. 

\section{Acknowledgement} 
This work is partly supported by the National Natural Science
Foundation of China under Grants No. 12247108 and
the China Postdoctoral Science Foundation under Grant
No. 2022M720360 and No. 2022M720359. This work is partly supported by the Spanish
Ministerio de Economia y Competitividad (MINECO)
and European FEDER funds under Contracts No.
FIS2017-84038- C2-1-P B, PID2020-112777GB-I00, and
by Generalitat Valenciana under contract PROME-TEO/2020/023. This project has received funding
from the European Union Horizon 2020 research and
innovation programme under the program H2020-
INFRAIA-2018-1, grant agreement No. 824093 of
the STRONG-2020 project. 
\bibliography{refs.bib}
\end{document}